\documentclass{article}

\usepackage{arxiv}

\usepackage[utf8]{inputenc} 
\usepackage[T1]{fontenc}    
\usepackage{hyperref}       
\usepackage{url}            
\usepackage{booktabs}       
\usepackage{amsfonts}       
\usepackage{nicefrac}       
\usepackage{microtype}      
\usepackage{lipsum}		
\usepackage{graphicx}
\usepackage[square]{natbib}
\usepackage{doi}
\usepackage{xcolor}
\usepackage{tcolorbox}
\usepackage{listings}

\colorlet{punct}{red!60!black}
\definecolor{background}{HTML}{EEEEEE}
\definecolor{delim}{RGB}{20,105,176}
\colorlet{numb}{magenta!60!black}

\lstdefinelanguage{json}{
    basicstyle=\normalfont\ttfamily,
    numbers=left,
    numberstyle=\scriptsize,
    stepnumber=1,
    numbersep=8pt,
    showstringspaces=false,
    breaklines=true,
    frame=lines,
    backgroundcolor=\color{background},
    moredelim=[is][\bfseries]{!!}{!!},
    literate=
     *{0}{{{\color{numb}0}}}{1}
      {1}{{{\color{numb}1}}}{1}
      {2}{{{\color{numb}2}}}{1}
      {3}{{{\color{numb}3}}}{1}
      {4}{{{\color{numb}4}}}{1}
      {5}{{{\color{numb}5}}}{1}
      {6}{{{\color{numb}6}}}{1}
      {7}{{{\color{numb}7}}}{1}
      {8}{{{\color{numb}8}}}{1}
      {9}{{{\color{numb}9}}}{1}
      {:}{{{\color{punct}{:}}}}{1}
      {,}{{{\color{punct}{,}}}}{1}
      {\{}{{{\color{delim}{\{}}}}{1}
      {\}}{{{\color{delim}{\}}}}}{1}
      {[}{{{\color{delim}{[}}}}{1}
      {]}{{{\color{delim}{]}}}}{1},
}

\title{Multi-Agent Systems Powered by Large Language Models: Applications in Swarm Intelligence}

\author{ \href{https://orcid.org/0000-0000-0000-0000}{\includegraphics[scale=0.06]{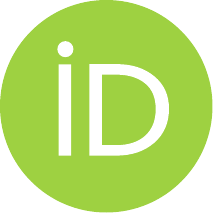}\hspace{1mm}Cristian~Jimenez-Romero} \\
	ETIS-Lab Faculty of Computer Science\\
	CY Cergy Paris University\\
	Cergy, France \\
	\texttt{cristian.jimenez-romero@cyu.fr} \\
	\And
	\href{https://orcid.org/0000-0001-8869-215X}{\includegraphics[scale=0.06]{orcid.pdf}\hspace{1mm}Alper~Yegenoglu} \\
	\texttt{alper.yegenoglu@rwth-aachen.de} \\
 	\And
	\href{https://orcid.org/0000-0002-1736-3559}{\includegraphics[scale=0.06]{orcid.pdf}\hspace{1mm}Christian~Blum} \\
	Artificial Intelligence Research Institute (IIIA-CSIC)\\
	Campus of the UAB\\
	Bellaterra, Spain \\
	\texttt{christian.blum@iiia.csic.es} \\
}

\date{}

\renewcommand{\shorttitle}{Integration of LLMs into MAS simulations}

\hypersetup{
pdftitle={A template for the arxiv style},
pdfsubject={q-bio.NC, q-bio.QM},
pdfauthor={Cristian Jimenez-Romero},
pdfkeywords={First keyword, Second keyword, More},
}

\begin{document}
\maketitle

\begin{abstract}

This work examines the integration of large language models (LLMs) into multi-agent simulations by replacing the hard-coded programs of agents with LLM-driven prompts. The proposed approach is showcased in the context of two examples of complex systems from the field of swarm intelligence: ant colony foraging and bird flocking. 
Central to this study is a toolchain that integrates LLMs with the NetLogo simulation platform, leveraging its Python extension to enable communication with GPT-4o via the OpenAI API. This toolchain facilitates prompt-driven behavior generation, allowing agents to respond adaptively to environmental data. For both example applications mentioned above, we employ both structured, rule-based prompts and autonomous, knowledge-driven prompts. Our work demonstrates how this toolchain enables LLMs to study self-organizing processes and induce emergent behaviors within multi-agent environments, paving the way for new approaches to exploring intelligent systems and modeling swarm intelligence inspired by natural phenomena. We provide the code, including simulation files and data at \url{https://github.com/crjimene/swarm_gpt}.

\end{abstract}


\section{Introduction}

\subsection{From Rule-Based to LLM-Driven Agents: A New Paradigm in ABMS}

In this study, we use the terms ``agent'' and ``multi-agent'' based on their foundational meanings in agent-based modeling and simulation (ABMS)~\citep{5429318}, while extending them through the integration of large language models (LLMs)~\citep{10.1145/3641289}. Traditionally, an agent in ABMS is an autonomous entity with localized decision-making abilities, interacting with its environment and other agents according to predefined rules or principles. A multi-agent system (MAS)~\citep{wooldridge2009introduction} refers to a collection of such agents operating within a shared environment, where global behaviors emerge from their local interactions. Examples of ABMS include simulations of ecosystem dynamics, urban development, and social interactions~\citep{heckbert2010agent,chen2012agent,bianchi2015agent}.

Beyond rule-based agents, ABMS literature also explores agents controlled by neural networks, especially in applications that require adaptability or learning. Techniques like reinforcement learning and evolutionary strategies are commonly used to optimize agent behaviors in dynamic environments~\citep{hecker2015beyond,ning2024survey,liu2024evolutionary}. 
Other research investigates the application of biologically inspired architectures, particularly spiking neural networks, to develop solutions that are both energy-efficient and computationally effective~\cite{fang2017achieving,putra2024snn4agents}. Within these systems, emergent behavior plays a pivotal role, facilitating the effective handling of complex tasks~\citep{jimenez2024emergent}.

Recently, these terms have been introduced in a different context within artificial intelligence. Here, AI agents often represent task-oriented entities designed to autonomously achieve specific goals, such as generating dialogues or managing workflows. These agents typically focus on individual task execution rather than the emergent dynamics central to ABMS (~\citep{talebirad2023multi,kannan2024smart,li2024survey}).

In our work, we employ the terms ``agent'' and ``multi-agent'' in the context of ABMS while integrating LLMs to guide agent behaviors. An agent in our simulations can be either LLM-steered or rule-based. We incorporate LLMs to guide agent behaviors in two ways: (1) simulations consisting entirely of LLM-steered agents, and (2) hybrid simulations where LLM-steered agents interact with traditional rule-based agents. This means our simulations can have populations of agents that are completely LLM-based or a mix of LLM-based and rule-based agents within the same environment.

We aim to explore the potential advantages of leveraging the decision-making and pattern-generation capabilities of LLMs to augment ABMS. Specifically, we are interested in investigating whether integrating LLMs can help us model emergent behavior using the language processing capabilities and the knowledge base of LLMs.

From this point forward, when we refer to ``agents,'' we mean agents within the ABMS framework that may incorporate LLM intelligence.

\subsection{Motivation} The field of agent-based simulations has rapidly evolved, driven by advances in artificial intelligence (AI) and computational power. These simulations, which model the interactions of autonomous agents within a defined environment, are increasingly being enhanced by the integration of generative AI, particularly LLMs. In this context, LLMs---with their capacity to process and generate human-like text---offer a novel means of guiding and influencing agent behaviors in real-time. A critical aspect of this integration is prompt engineering, which is the careful design of prompts that serve as instructions for the agents, dictating how they should respond to their environment.

The motivation and contribution of this work are found in the presentation of a toolchain that integrates LLMs with agent-based simulations within the NetLogo environment~\citep{tisue2004netlogo,amblard2015introduction}, a platform widely recognized in the complexity science community for its robustness and versatility. NetLogo's value as an educational tool spanning various academic levels further underscores its importance, making it an ideal choice for demonstrating the integration of advanced AI methods into multi-agent simulations. 

Our study investigates two distinct approaches to utilizing LLMs within multi-agent environments, focusing on the role of prompt engineering in shaping agent behavior. The first approach employs detailed, structured prompts within an ant colony simulation. These prompts are designed to specify behaviors under clearly defined conditions, such as following pheromone trails or retrieving food. This method allows for precise control over agent actions, enabling a rule-based system where each agent’s behavior is explicitly dictated by the LLM-generated instructions.

In contrast, The second approach explores the use of less structured, principle-based prompts in a bird flocking simulation. Here, the prompts rely on the LLM’s inherent understanding of complex concepts such as flocking dynamics and self-organization. Instead of requiring explicit, rule-based instructions, these prompts allow the LLM to handle the intricate behavioral patterns that would otherwise need numerous rules to define. This approach leverages the LLM’s capacity to intuitively model these dynamics, enabling behaviors that emerge naturally from agents' interactions with each other and their environment. As we will show, the LLM can produce adequate and adaptive agent behaviors that realistically reflect complex, emergent patterns within the simulation.

\subsection{Research Objectives}
The following are the main objectives of our research:
\begin{itemize}
	\item To assess the efficacy of structured prompts in the context of the rule-based ant colony foraging simulation of NetLogo, which is a classic MAS model that demonstrates swarm intelligence principles based on how real ants find food and communicate via pheromones. It is widely used in artificial intelligence, complexity science, and optimization research.

	\item To assess the efficacy of structured prompts in NetLogo's knowledge-driven bird flocking simulation, which is also a classic model demonstrating self-organizing behavior in MAS. It is inspired by Craig Reynolds' ``Boids'' model~\citep{reynolds1987flocks}, which simulates how birds, fish, or other animals move in cohesive groups without a central leader.
    
	\item To present a comprehensive toolchain that combines LLMs with multi-agent simulation environments, offering a new method for modeling and analyzing swarm behavior in complex systems.
\end{itemize}

This investigation aims to explore how LLMs, through effective prompt engineering, can be integrated into multi-agent systems to model and guide emergent, self-organizing behaviors in simulated environments.

\subsection{Background and Related Work}

The integration of generative AI into multi-agent systems represents a burgeoning field that seeks to enhance the autonomy, adaptability, and realism of agent behaviors in simulations. This approach leverages the vast knowledge embedded within LLMs to influence agent interactions in ways that were previously unfeasible with traditional rule-based systems. The use of generative AI in multi-agent simulations has opened new avenues for exploring complex behaviors, emergent dynamics, and adaptive systems.

In particular, the integration of LLMs with agent-based simulations represents a significant convergence of natural language processing (NLP) and complex systems modeling. LLMs, with their ability to generate human-like text and understand complex linguistic patterns, have transformed various fields within artificial intelligence, particularly in automating and interpreting language-based tasks. On the other hand, agent-based simulations are a robust framework for modeling complex systems where individual agents interact with each other and their environment, potentially leading to emergent behaviors. The use of LLMs in simulations may hereby vary widely, from highly structured, rule-based prompts that delineate specific actions to more generalized prompts that rely on the LLM’s broader knowledge base. This study highlights two distinct methodologies in applying LLM capabilities to simulate emergent, multi-agent behaviors with varying degrees of prompt specificity and autonomy.

Integrating LLMs with agent-based simulations presents transformative opportunities across various domains, enhancing the realism and complexity of simulations. This integration can significantly improve the modeling of social systems, industrial automation, and multi-agent interactions.

\cite{park2023generative} introduce an LLM-driven agent that can engage and converse with both humans and other AI agents. The agent has the capability to generate text that can be comprehended and interpreted by other agents. This facilitates clear communication between them, fostering effective interactions and collaboration. The simulated environment functions as a sandbox composed of text, allowing the agent to perceive and interpret the surrounding context effectively. This setting enables the agent to navigate and interact with the provided information. Inspired by the work of~\cite{park2023generative}, \cite{junprung2023exploring} presents two LLM-driven experiments, two-agent negotiation, and a six-agent murder mystery game to simulate human behavior. The author describes the behavior of three categorical different LLM-driven simulations and discusses the limitations of large-scale language models.

\cite{gao2023} create a framework for social network simulation called $S^{3}$. They simulate motion, attitude, and interactive behaviors to emulate social behavior. Due to the changing environment, the agents have to adapt and retain a memory to utilize past experiences and adjust their behavior. They observe the emergence of collective behavior among the agents and conclude their environment holds potential for further exploration in the fields of social sciences and informed decision-making. This insight suggests that the dynamics observed could provide valuable perspectives on group interactions and collaborative processes.

The research of~\cite{dasgupta2023collaborating} investigates the use of LLMs to improve the decision-making abilities of AI agents that interact with their environment. The proposed system consists of three parts: a Planner that uses a pre-trained LLM to generate instructions, a reinforcement-learning agent, the Actor, that carries out these instructions, and a Reporter that provides environmental feedback to the Planner. The Planner reads a description of the task and breaks it down into simple instructions for the Actor, who was trained to understand simple instructions and operates upon them. The Reporter observes the effects of the Actor's actions on the environment and communicates this information in a text-based form back to the Planner.       
The system is tested on complex tasks that require reasoning and gathering information, and the results show that it outperforms traditional reinforcement learning methods, especially when using larger language models. The researchers demonstrate that Large language models (70 billion parameters) consistently outperformed smaller language models (7 billion parameters) in the experiments, indicating that larger models have resilience against noisy or irrelevant information and greater capacity for the complex reasoning required to solve these tasks. 
\cite{zhu2023ghost} present Ghost in the Minecraft (GITM), a framework for developing general capable agents in the Minecraft world. In contrast to previous approaches, especially reinforcement learning algorithms, GITM uses large language models to achieve high success rates, e.g. in the "Obtain Diamond" task. Typical reinforcement learning-based agents often struggle with the complexity of Minecraft due to the long time horizon of the task, which can lead to difficulties in learning and adapting. In contrast, \cite{zhu2023ghost} leverages LLMs to enable a hierarchical decomposition of complex tasks into manageable sub-goals and structured actions. This approach yields significantly higher efficiency and robustness, allowing agents to better navigate and interact with the Minecraft environment. GITM integrates the logical reasoning and knowledge base of LLMs with text-based knowledge and memory, enabling effective interaction with the environment and the pursuit of intricate, long-term objectives. The article demonstrates the potential of LLMs for the development of generally capable agents in open, complex environments.

Recently, researchers incorporated LLM into swarm systems to leverage the reasoning and knowledge capabilities of these models~
\citep{gao2024large,qu2024federated}. \cite{strobel2024llm2swarm} integrate LLMs into robot swarms to enhance their reasoning, planning, and collaboration abilities. They exchange the robot programming controller by proposing two changes: 1) An indirect integration uses LLMs to generate and validate the programming of the controller before or during the deployment. This approach improves efficiency and reduces human error by automating the design process. 2) A direct integration implements a separate LLM for each robot during deployment, enabling the robot to plan, reason, and collaborate using natural language. The LLM-driven robots can detect and respond to unexpected behaviors and are more resilient in dynamic environments without prior information.

\cite{feng2024model} present an algorithm aimed at adapting LLM experts using collaborative search techniques inspired by swarm intelligence. This method allows several LLMs to collaborate in exploring the weight space to optimize a specific utility function without the need for extensive fine-tuning data or strong assumptions about the models involved. In their work, each LLM can be treated as a particle within a swarm navigating within the weight space and adjusting its position based on its best or worst-found solutions. The algorithm demonstrates flexibility in different single or multi-task objectives. Due to their collaborative search approach the LLM experts can discover unseen capabilities, which enables the transition from weak to strong performance levels. 

In their work, called Swarm-GPT, \cite{jiao2023swarm} integrate LLMs with motion-based planning to automate Unmanned Aerial Vehicle (UAVs) swarm choreography. Users are able to generate synchronized drone performance via language commands. Swarm-GPT is able to utilize LLMs to create UAVs formations and movements which are synchronized to music. The system includes a trajectory planner that utilizes waypoints generated by the LLM, guaranteeing that the drone movements are both collision-free and feasible. Swarm-GPT has been effectively demonstrated at live events, highlighting its practical application and ability to perform in real time.

\cite{liu2024multi} explore the application of multimodal LLMs to control the formation of UAVs using image and text inputs. The researchers first pre-trained an LLM on a single UAV, demonstrating the LLM's potential to interpret and execute commands effectively. Then they expanded their approach to coordinate multiple UAVs in formation. The multimodal LLM recognizes environmental signals from the images captured by the primary drone (via a camera). Then, the pre-trained LLM analyzes the data and generates instructions for managing the UAV to attain a specified formation.

Another application in language-guided formation control is presented by~\cite{liu2024language}. The authors propose a framework called Language-Guided Pattern Formation (LGPF) for swarm robotics. Their system employs an LLM to translate a high-level pattern description into specific actions for a swarm of robots, integrating multi-agent reinforcement learning for detailed control. The LGPF framework allows for intuitive and flexible control of robot swarms, enabling them to achieve complex formations guided by natural language instructions. 

\section{Methods}
In this study, we employed two distinct simulations of the behavior of social insects to explore the integration of LLMs in guiding agent behaviors within multi-agent systems. The experiments were designed to investigate the effectiveness of structured, rule-based prompts in one scenario and principle-based, knowledge-driven prompts in the other one. Both simulations utilize the LLM to process environmental inputs and generate agent actions, providing insights into how LLMs can be leveraged to model complex behaviors such as foraging and flocking.

\textbf{Structured rule-based} prompts are designed with explicit, predefined instructions that guide the LLM to generate deterministic agent actions. These prompts specify exact conditions and responses, ensuring consistent and predictable agent behaviors. For example, in a foraging scenario, structured prompts might include direct rules for following pheromone trails or picking up food when encountered.

\textbf{Knowledge-driven} prompts, on the other hand, rely on the LLM's inherent understanding of broader behavioral concepts and principles. These prompts are less rigid and provide the LLM with general guidelines, enabling more adaptive and flexible agent behaviors. In the context of a bird flocking simulation, such prompts might encourage behaviors based on principles like alignment, cohesion, and separation without specifying exact actions, allowing the LLM to synthesize responses that foster emergent, self-organizing dynamics.

\subsection{Toolchain for LLM-Driven Multi-Agent Simulations with NetLogo}

\begin{figure}
    \centering
    \includegraphics[width=0.95\linewidth]{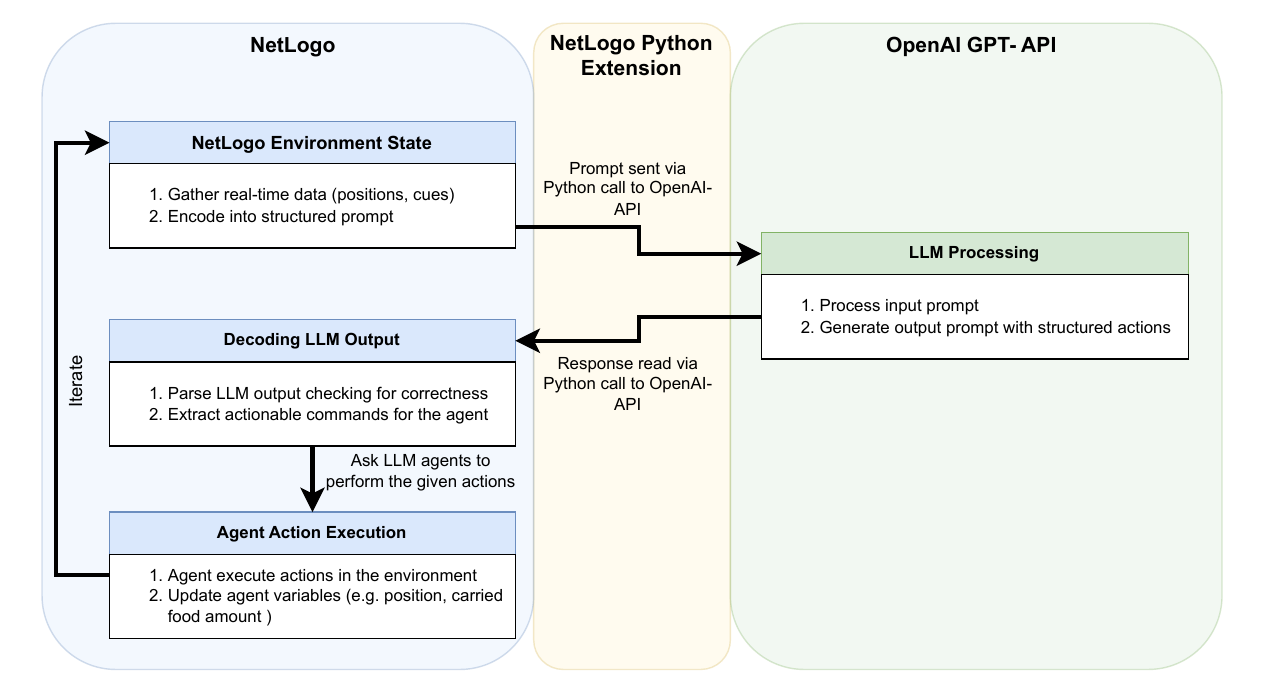}
    \caption{Diagram illustrating the toolchain for LLM-driven multi-agent simulations, integrating NetLogo and GPT-4o via the Python extension and OpenAI API. The workflow showcases a closed-loop process where environmental states are encoded into structured prompts, processed by GPT-4o to generate behavior suggestions, decoded into executable actions, and iteratively executed by agents within the NetLogo simulation environment.}
    \label{fig:netlogo_gpt_framework}
\end{figure}

Figure \ref{fig:netlogo_gpt_framework} illustrates the toolchain for LLM-driven multi-agent simulations with NetLogo, showing the integration between NetLogo, GPT-4o, and the Python extension. The following describes each step of the workflow:

\begin{enumerate}
\item \textbf{Environment Encoding:} The simulation toolchain leverages NetLogo to capture real-time environmental states, including agent positions, inter-agent interactions, and other relevant environmental variables depending on the simulation (e.g. pheromone concentrations). These data are encoded into structured prompts that convey a comprehensive environmental context to the LLM. This encoding ensures that the LLM receives timely, accurate input representing dynamic changes in the environment.

\item \textbf{Python Extension Integration:} NetLogo uses its Python extension to facilitate communication with GPT-4o via the OpenAI API. This extension allows NetLogo to send encoded environmental data as prompts to the LLM and receive structured responses, enabling the interaction between the simulation platform and the LLM.

\item \textbf{LLM Processing:} The structured prompts are processed by GPT-4o, which interprets the input data and generates agent behavior suggestions based on encoded environmental information. The LLM’s ability to process complex, context-rich data allows it to infer and propose actions that adhere to predefined rules (for structured prompts) or leverage general behavioral principles (in principle-based prompts). This stage ensures that agent responses align with the overall objectives of the simulation, be it foraging success or cohesive flocking.

\item \textbf{Decoding LLM Output:} The LLM output, formatted as a structured JSON or Python dictionary, is translated into executable actions predefined within the NetLogo simulation. This step converts the structured actions generated by the LLM into precise instructions for agents, such as movement vectors, state transitions, or pheromone release behaviors. The Python extension facilitates this process by receiving the LLM responses from the OpenAI API and converting them into a NetLogo-compatible data structure. This translation mechanism ensures both syntactic and semantic alignment between the LLM’s output and the data format required by the simulation.

\item \textbf{Agent Action Execution and Iterative Process:} The decoded commands are executed by the agents in NetLogo, updating their states and behaviors in response to the LLM's instructions. This action directly modifies the simulation environment, forming a closed-loop system where each action feeds back into the environmental context for the next iteration. The iterative process ensures that agent behaviors continuously respond to evolving environmental conditions and LLM feedback, fostering emergent behaviors and adaptive responses.
\end{enumerate}
The following sections detail the setup, LLM integration, and procedures used in each experiment.

\section{Experiment 1: Ant Colony Foraging Simulation}
\begin{figure}[t]
    \centering
    \includegraphics[width=0.7\linewidth]{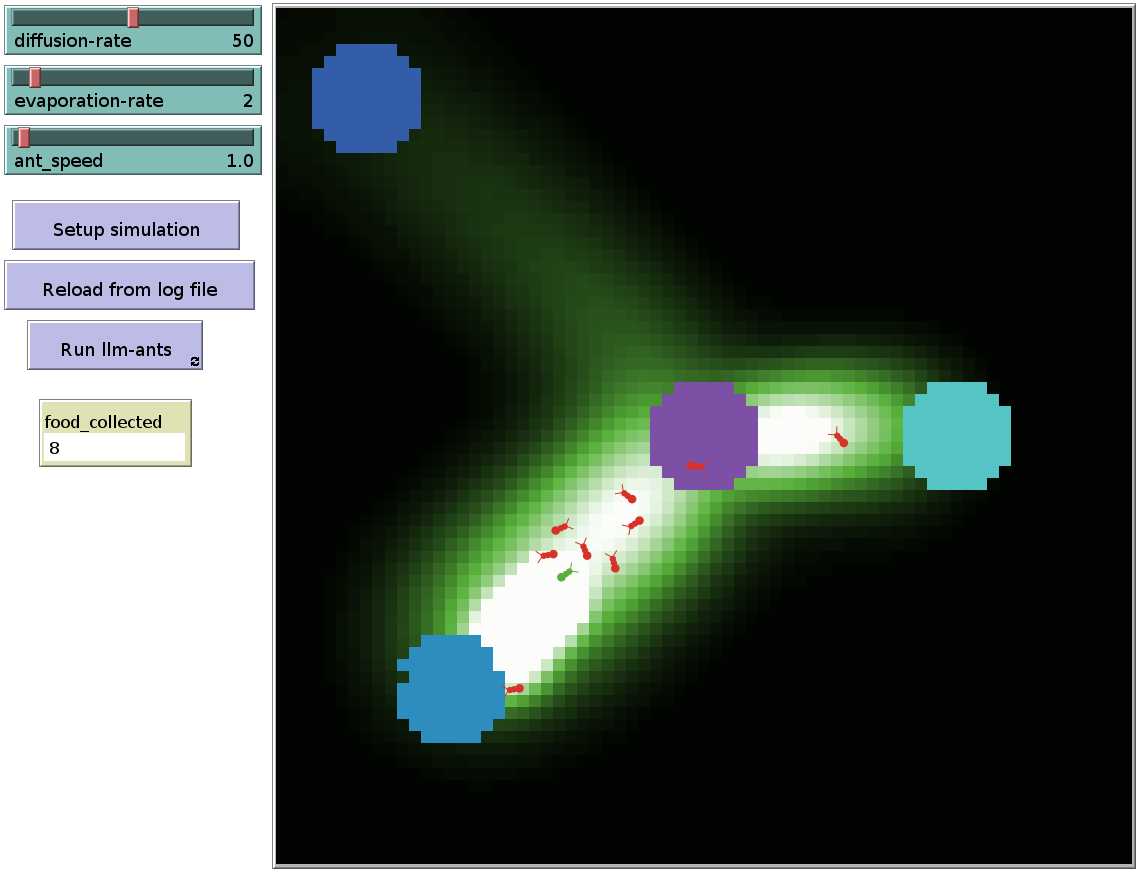}
    \caption{Ant foraging simulation in NetLogo. The central circle depicts the nest area, while the three blue circles nearby indicate food sources. }
    \label{fig:ant_colony_interface}
\end{figure}

As mentioned above, this experiment is based on the ant foraging model implemented in the NetLogo library (see~\url{https://ccl.northwestern.edu/netlogo/models/Ants}). It takes place in a two-dimensional foraging area consisting of designated food sources scattered throughout the environment and a central nest where the ants must return to deposit the food they collect. The environment is designed to mimic natural foraging conditions, where agents (ants) must navigate to find food and return it to the nest while interacting with environmental cues such as pheromone trails and nest scents; see Figure~\ref{fig:ant_colony_interface}.

\begin{itemize}
\item \textbf{Agents}: The simulation features stateless ants, each operating as an independent agent without memory of past actions. These ants rely entirely on real-time environmental inputs and LLM-generated prompts to determine their behaviors. The agents are designed to follow explicit, rule-based instructions derived from the LLM, ensuring that their actions are predictable and consistent with predefined conditions.
\item \textbf{LLM Integration}: OpenAI GPT-4o is employed to process structured prompts that define the ants' behaviors. The default API parameters are used, with the exception of the temperature, which is set to 0.0 to ensure deterministic results based on the provided inputs. The LLM receives real-time environmental information and generates actions according to a predefined set of rules. These structured prompts ensure that the ants' responses are clearly defined and predictable, enabling systematic analysis of their behavior. Nevertheless, there is still a small chance that the LLM may occasionally generate responses that deviate from the specified rules. 
\end{itemize}

\subsection{Procedure}
\subsubsection{Prompt Design}
The prompt is structured as a zero-shot prompt, requiring the LLM to generate accurate responses without relying on prior examples or contextual memory from previous interactions. This intentional design maintains stateless prompts to control the agents. We employed an API function that does not retain conversation context between prompts, making each interaction independent and requiring the LLM to interpret and respond solely based on the current input. The final prompt used in our experiments resulted from several iterations in a trial-and-error process to ensure the LLM could effectively understand the environment and rules and execute the expected tasks accordingly. For this experiment, precise behavioral rules were provided for the ants within the simulation, including instructions such as following pheromone trails when searching for food, picking up food when encountered, and releasing pheromones to mark food sources.

\subsubsection{Tuning Process}

Our initial approach utilized minimal instructions, providing a general description of the foraging task to assess how effectively ants could perform without specific guidance. This method aimed to determine the necessity of detailed instructions for reproducing pheromone-guided foraging behavior.

As the project progressed, it became clear that more explicit rules were essential for consistent and realistic ant behavior. Throughout this iterative process, the language model offered valuable feedback by highlighting misunderstandings or ambiguities in the prompts. This feedback was instrumental in refining the prompts to enhance the ants' performance. Below, we present an analysis of how these prompts evolved, focusing on specific improvements and the reasoning behind each iteration.

\textbf{Iteration 1}

\begin{itemize}
\item Prompt Text

\begin{lstlisting}[language=json,firstnumber=1,basicstyle=\small]
You are an ant in a 2D simulation tasked with finding food, marking the path to food with trails of pheromones, and using nest scent to navigate back to the nest when carrying food.

Format your actions as a Python dictionary with these keys and options:
    "move-forward": True or False,
    "rotate": "left", "right", or "none",
    "pick-up-food": True or False,
    "drop-pheromone": True or False,
    "drop-food": True or False.

You will be provided with environment information. Keep your response concise, under 35 tokens.

Current environment:
    -Pheromone concentration (Left: 0, Front: 0, Right: 0),
    -Nest presence: True,
    -Nest scent (Left: 196.84, Front: 196.39, Right: 195.76),
    -Food concentration at your location: 0,
    -Carrying food status: True
\end{lstlisting}

\item \textbf{Observed Behavior}

In this first attempt, we provided general instructions to establish a baseline for ant behavior. The ants were tasked with finding food, marking paths with pheromones, and using nest scent to return home when carrying food. However, simulations revealed inconsistencies. Ants often failed to follow pheromone and nest scent gradients effectively, sometimes moving away from stronger cues. Some ants released pheromones unexpectedly, while others exhibited random movement patterns. This inconsistency prevented the colony from displaying an organized foraging behavior, indicating that clearer rules were needed for actions such as pheromone release and gradient follow-up.

\item \textbf{Analysis}

While this prompt established the basic framework for the simulation, it lacked specific guidance on how ants should interpret and prioritize environmental cues or resolve conflicting signals. The absence of detailed instructions led to ambiguous behaviors, including inconsistencies in following pheromone and nest scent gradients. This highlighted the need for more explicit rules to ensure consistent and organized swarm behavior.
\end{itemize}

\textbf{Iteration 2}
\begin{itemize}
\item \textbf{Prompt Text}

We add an instruction to the prompt to prioritize nest scent over pheromone trails when carrying food.

\begin{lstlisting}[language=json,firstnumber=1,basicstyle=\small]
You are an ant in a 2D simulation tasked with finding food, marking the path to food with pheromone trails, and using nest scent to navigate back to the nest when carrying food. !!Prioritize nest scent over pheromone trails when carrying food!!.
...
\end{lstlisting}

\item \textbf{Observed Behavior}

To address the issues from the first prompt, we added a directive for ants to prioritize nest scent over pheromone trails when carrying food, aiming to better mimic foraging ant behavior. Despite this improvement, ants still exhibited inconsistencies in following pheromone and nest scent gradients. When nest scent and pheromone trails had similar strengths, ants demonstrated conflicting actions. Additionally, the prompt did not specify behaviors for ants not carrying food, leading to inefficient exploration. Ants tended to rotate aimlessly near the nest and were slow to venture outward, showing the need for clearer guidance to improve exploration efficiency.

\item \textbf{Analysis}\textit{}

Introducing prioritization helped align the ants' actions when carrying food, but inconsistencies in following scent gradients persisted. Ants not carrying food and not sensing any pheromones tended to remain near the nest without effectively exploring the environment. This emphasized the necessity for comprehensive guidance covering all possible states and clearer instructions on responding to environmental cues to enhance exploration efficiency.
\end{itemize}

\textbf{Iteration 3}
\begin{itemize}
\item \textbf{Prompt Text}

As before, with added clarifications in the current environment:

\begin{lstlisting}[language=json,firstnumber=1,basicstyle=\small]
...
    -Nest presence: True !!(You are currently at the nest)!!,
    -Carrying food status: True !!(You are currently carrying food)!!
...
\end{lstlisting}

\item \textbf{Observed Behavior}

We observed that ants sometimes failed to pick up food or drop it at the nest, possibly due to a lack of awareness of their current state. To rectify this, we explicitly stated their status in the prompt, such as whether they were at the nest or carrying food. This redundancy ensured that ants performed correct actions in these situations. However, inconsistencies in following pheromone and nest scent gradients remained. Ants continued to exhibit limited exploration when not carrying food, tending to stay near the nest rather than venturing into new areas or effectively following pheromone trails.

\item \textbf{Analysis}

Explicitly stating the ants' status improved decision-making by providing clear context, leading to better execution of actions like picking up and dropping food. Yet, the lack of specific instructions on following scent gradients meant ants still showed inconsistencies in navigating toward pheromone trails or nest scent. Their inefficient exploration highlighted the need for clearer guidance to enhance movement away from the nest.
\end{itemize}

\textbf{Iteration 4}
\begin{itemize}
\item \textbf{Prompt Text} 

We add an instruction to the prompt to use the highest pheromone scent to navigate to food when not carrying any.
\begin{lstlisting}[language=json,firstnumber=1,basicstyle=\small]
You are an ant in a 2D simulation. Your task is to pick up food and release it at the nest. Use nest scent to navigate back to the nest when carrying food, prioritizing nest scent over pheromones. !!Use highest pheromone scent to navigate to food when not carrying any!!.
...
\end{lstlisting}

\item \textbf{Observed Behavior}

To guide ants not carrying food, we specified that they should navigate toward food using the highest pheromone concentration. Their ability to find food sources when pheromone trails were present was clearly improved in this way. However, inconsistencies in following pheromone gradients persisted. In the absence of pheromones or nest scents, ants tended to remain near the nest, exhibiting inefficient exploration behaviors.

\item \textbf{Analysis}

By distinguishing between the states of carrying and not carrying food, we enhanced the ants' foraging efficiency when environmental cues were available. Nonetheless, inconsistencies in following pheromone gradients indicated that ants needed clearer instructions on interpreting and acting upon varying scent intensities. The lack of an effective exploration strategy, when cues were absent, remained a challenge.
\end{itemize}

\textbf{Iteration 5}
\begin{itemize}
\item \textbf{Prompt Text}

Environmental information about pheromone concentration and nest scent presented with directional cues instead of quantities:
\begin{lstlisting}[language=json,firstnumber=1,basicstyle=\small]
...
Current environment:
    -Higher Pheromone Concentration: !!Front!!,
    -Nest Presence: False (You are not currently at the nest),
    -Stronger Nest Scent: !!Left!!,
    -Food Concentration at your location: 0,
    -Carrying Food Status: True (You are currently carrying food)}
\end{lstlisting}
    
\item \textbf{Observed Behavior}

Recognizing the need for better interpretation of environmental cues, we modified how information was presented by using directional descriptions instead of numerical values—--e.g., ``Higher Pheromone Concentration: Front'' and ``Stronger Nest Scent: Left.'' This adjustment significantly improved the ants' ability to follow pheromone and nest scent gradients. Ants became more consistent in moving toward stronger cues, enhancing their navigation and foraging efficiency.

However, when no scents were detected, ants still showed limited exploration, often remaining near the nest rather than actively searching new areas. This indicated that while gradient following had improved, the exploration strategy was still inefficient in the absence of sensory cues.

\item \textbf{Analysis}

Using directional cues provided clearer guidance on responding to environmental gradients, resolving many inconsistencies observed in previous prompts. From Prompt 5 onward, ants became more adept at following pheromone and nest scent gradients, leading to a more organized foraging behavior. Despite these improvements, ants' exploration remained inefficient when no sensory cues were present, indicating a need for further instructions to promote effective exploration.
\end{itemize}

\textbf{Iteration 6}
\begin{itemize}
\item \textbf{Prompt Text}

We add an instruction to the prompt to release pheromones on food sources and while carrying food.

\begin{lstlisting}[language=json,firstnumber=1,basicstyle=\small]
You are an ant in a 2D simulation. Your task is to pick up food and release it at the nest. !!Release pheromone on food source and while you are carrying food.!! Use nest scent to navigate back to the nest when carrying food, prioritizing nest scent over pheromones. Use highest pheromone scent to navigate to food when not carrying any.
...
\end{lstlisting}

\item \textbf{Observed Behavior}

To encourage trail formation back to the nest, we instructed ants to release pheromones while carrying food. This led to stronger trails and improved the efficiency of other ants in locating food sources. With the improved gradient-following ability from Prompt~5, ants were more consistent in navigation.

Nevertheless, in the absence of pheromones and nest scents, ants still exhibited limited exploration behaviors, tending to stay near the nest. This indicated that their exploration strategy was still inefficient and required refinement.

\item \textbf{Analysis}

By enhancing pheromone deposition during food transport and improving gradient following, we boosted colony cooperation and foraging success. However, the persistent issue of limited exploration in scent-free areas indicated that additional instructions were necessary to promote outward movement and enhance exploration efficiency.
\end{itemize}

\textbf{Iteration 7}
\begin{itemize}
\item \textbf{Prompt Text}

We added the word ``only'' in the prompt as follows: 

\begin{lstlisting}[language=json,firstnumber=1,basicstyle=\small]
You are an ant in a 2D simulation. Your task is to pick up food and release it at the nest. Release pheromone on food source and while you are carrying food. Use nest scent to navigate back to the nest !!only!! when carrying food, prioritizing nest scent over pheromones. Use highest pheromone scent to navigate to food when not carrying any.
...
\end{lstlisting}

\item \textbf{Observed Behavior}

In earlier iterations, ants sometimes prioritized nest scent over pheromones even when not carrying food, leading them to return to the nest unnecessarily. With this clarification, the ants began to prioritize the nest scent appropriately, using it only when they were carrying food. However, ants still exhibited limited exploration when no sensory cues were present, tending to remain near the nest rather than actively searching new areas.

\item \textbf{Analysis}

Adding ``only'' to the instruction text was crucial to ensure that the ants did not prioritize the scent of the nest when they were looking for food. This eliminated unnecessary returns and improved foraging efficiency.
\end{itemize}

\textbf{Iteration 8}
\begin{itemize}
\item \textbf{Prompt Text}

We added an instruction to the prompt to move away from the nest and rotate randomly if not carrying any food and not sensing any pheromone.

\begin{lstlisting}[language=json,firstnumber=1,basicstyle=\small]
You are an ant in a 2D simulation. Your task is to pick up food and release it at the nest. Release pheromone on food source and while you are carrying food. Use nest scent to navigate back to the nest only when carrying food, prioritizing nest scent over pheromones. Use highest pheromone scent to navigate to food when not carrying any. !!Move away from nest and rotate randomly if you are not carrying any food and you are not sensing any pheromone!!.
...
\end{lstlisting}

\item \textbf{Observed Behavior}

In previous iterations, we observed limited exploratory behavior of the ants in areas without scents. To address this, we introduced a directive for proactive exploration. This approach improved exploration, with ants venturing further from the nest and discovering food sources in fewer simulation steps. However, a noticeable bias concerning the rotation remained, particularly around the nest, indicating that the randomness was not functioning as efficiently as intended.

\item \textbf{Analysis}

By instructing ants to move away from the nest and rotate randomly when not carrying food and not sensing pheromones, we encouraged them to explore new areas more effectively. This change increased the likelihood of ants finding food, as they ventured further from the nest rather than lingering nearby.
\end{itemize}

\textbf{Iteration 9}
\begin{itemize}
\item \textbf{Prompt Text}

We expanded the rotation options to include ``random'':

 \begin{lstlisting}[language=json,firstnumber=1,basicstyle=\small]
...
   "rotate": "left", "right", "none", !!or "random"!!
...
\end{lstlisting}

\item \textbf{Observed Behavior}

With this adjustment, ants demonstrated more varied and unpredictable movement patterns during exploration. They effectively moved away from the nest and searched a wider area, increasing their chances of encountering food sources more quickly and efficiently.

\item \textbf{Analysis}

To enhance the randomness of the ants' exploration, we expanded their rotation options to include ``random.'' This meant that when the LLM selected ``random'' as the rotation action, it was making a high-level decision to delegate the choice of direction to chance. In the simulation, this ``random' '  option was implemented at a programming level in NetLogo to randomly choose the direction of rotation either left or right.
\end{itemize}

Through iterative tuning, we significantly enhanced the simulated ants' behavior, making it more consistent with the ant foraging dynamics observed in the rule-based NetLogo model. Each prompt iteration addressed specific issues identified in simulations, with language model feedback guiding some of the adjustments. 

\subsubsection{Prompt Deployment}
The prompts are presented in a format that the LLM can process and output as a series of actionable commands. Communication with the LLM is facilitated through the OpenAI API, specifically using the \texttt{chat.completions} mechanism, which allows context-free messages to be passed at each step. This setup involves sending a system prompt that outlines the overall task and rules, followed by user prompts that provide real-time environmental information.

At each simulation step, NetLogo translates the agent's perception of its environment into the input variables described in the prompt. This ensures that the LLM has an accurate and up-to-date representation of the environment on which its decisions can be based. The LLM then generates a response formatted as a Python dictionary, containing specific actions the agent should take. The following example prompt illustrates the process:

\begin{itemize}
\item \textbf{System Prompt}

\begin{lstlisting}[language=json,firstnumber=1]
You are an ant in a 2D simulation. Your task is to pick up food and release it at the nest. Release pheromone on food source and while you are carrying food. Use nest scent to navigate back to the nest only when carrying food, prioritizing nest scent over pheromones. Use highest pheromone scent to navigate to food when not carrying any. Move away from nest and rotate randomly if you are not carrying any food and you are not sensing any pheromone. Format your actions as a Python dictionary with these keys and options: 

   "move-forward" (options: True, False)
   "rotate" (options: "left", "right", "none", "random" )
   "pick-up-food" (options: True, False)
   "drop-pheromone" (options: True, False)
   "drop-food" (options: True, False). 
   
You will be provided with environment information. Keep your response concise, under 45 tokens.
\end{lstlisting}

\item \textbf{Possible User Prompt}

\begin{lstlisting}[language=json,firstnumber=1]
This is your current environment: 
   -Highest Pheromone Concentration: None,
   -Nest Presence: True (You are currently at the nest),
   -Stronger Nest Scent: Front,
   -Food Concentration at your location: 0,
   -Carrying Food Status: False (You are not currently carrying food).
\end{lstlisting}

\item \textbf{Possible response from the LLM}
\begin{lstlisting}[language=json,firstnumber=1]
{
   "move-forward": True, 
   "rotate": "none", 
   "pick-up-food": False,
   "drop-pheromone": False,
   "drop-food": False
}
\end{lstlisting}
\end{itemize}

\subsection{System Monitoring}

The simulation monitors the ants’ efficiency in navigating the environment, specifically their ability to locate food, follow pheromone trails, and return food to the nest. Key metrics include the time taken to find and return food, the consistency of pheromone trail usage, and the accuracy of following the nest scent when carrying food. The observed behaviours and performance analysis are documented in the result section.

\section{Experiment 2: Bird Flocking Simulation}

As mentioned before, the bird flocking model of NetLogo (see \url{https://ccl.northwestern.edu/netlogo/models/Flocking}) is an implementation of the famous Boids model from~\citep{reynolds1987flocks}. The simulation takes place in two-dimensional airspace. Although this environment is relatively simple, it effectively replicates key flocking behaviors like group cohesion, allowing for the observation of flocking dynamics under varying conditions. By adjusting specific parameters, the simulation provides insights into how changes in the environment influence flocking behavior.
\begin{itemize}
\item \textbf{Agents}: The agents in this simulation are modeled as birds, each operating under principle-based prompts. Unlike rule-based systems, these birds are guided by general principles of flocking dynamics, that is, by alignment, separation, and cohesion~\citep{reynolds1987flocks}. These principles help the birds navigate their environment by adjusting their headings in response to the positions and headings of neighboring birds.
\item \textbf{LLM Integration}: The prompts provided to the LLM leverage its inherent knowledge of flocking dynamics, requiring it to apply these general principles to guide the behavior of each bird. The LLM is responsible for interpreting environmental data and generating responses that ensure the birds align with their flockmates, maintain an appropriate distance to avoid collisions, and stay cohesive as a group.
\end{itemize}

\subsection{Procedure}

\subsubsection{Prompt Design}

Similar to the setup in the case of ant foraging, prompts for the flocking task are structured as zero-shot prompts, meaning they operate without prior examples or contextual memory from previous interactions. The final prompt was tuned through several iterations (see below) in a trial-and-error process to ensure the LLM could effectively interpret the environment and calculate heading directions according to flocking principles. Each prompt guiding a bird is designed to determine its heading based on the three core principles of flocking dynamics as implemented in the NetLogo library: Separation (steering to avoid crowding neighbors), Alignment (steering towards the average heading of nearby birds), and Cohesion (steering towards the average position of nearby flockmates).

\subsubsection{Tuning Process}

As will be shown below, it was crucial to explicitly state in the prompt that the compass convention is used in the simulation. This alignment with NetLogo's world representation, where headings are measured in degrees---0 degrees pointing north, 90 degrees east, 180 degrees south, and 270 degrees west---was essential. Clearly defining this convention ensured that the LLM could accurately compute and adjust the birds' headings according to flocking dynamics, maintaining consistency in the agents' behavior within NetLogo’s simulation environment.

\textbf{Iteration 1}

\begin{itemize}
\item Prompt Text

\begin{lstlisting}[language=json,firstnumber=1,basicstyle=\small]

You are an agent in a 2D simulation. Your task is to determine your new heading based on the flocking principles of separation turn, alignment turn (average heading of neighbors), and coherence turn (average heading towards flockmates). The parameters for these principles are: maximum-separate-turn, maximum-align-turn, maximum-cohere-turn, minimum-separation-distance. The simulation provides the following information: Current heading, Neighbors in vision radius.

Provide your final new heading after applying these rules, expressed as an angle in degrees. The result should be in JSON format, with the key and value: "new-heading" (value: heading in degrees). Summarize your answer in no more than 120 words.

These are the flocking parameters:

    Maximum separate turn: 1.5
    Maximum align turn: 5
    Maximum cohere turn: 3
    Minimum separation: 1

This is your current environment:

    Current heading: 138 deg
    Neighbors in vision radius: neighbor_1: x: 0.53, y: -3.69, heading: 248 deg
\end{lstlisting}

\item \textbf{Observed Behavior}

In this initial attempt, we provided general instructions to establish a baseline for flocking behavior. The agents were instructed to determine their heading based on the principles of separation, alignment, and coherence. However, most of the LLM-generated responses were not interpretable by the simulation, as they did not adhere to the expected format. Additionally, even when successfully parsed, inconsistencies in the agents’ behavior were observed, preventing the emergence of flocking.

\item \textbf{Analysis}

While this prompt defined the basic framework for the simulation, it lacked constraints to enforce a structured response. In many cases, the LLM's output included extended textual and mathematical explanations before or alongside the JSON object, which interfered with proper parsing.
\end{itemize}

\textbf{Iteration 2}
\begin{itemize}
\item \textbf{Prompt Text}

An explicit instruction was added to limit the response to the JSON object only.

\begin{lstlisting}[language=json,firstnumber=1,basicstyle=\small]
...
Provide your final new heading after applying these rules, expressed as an angle in degrees. The result should be in JSON format !!only!!, with the key and value: "new-heading" (value: heading in degrees). Summarize your answer in no more than 120 words.
...
\end{lstlisting}

\item \textbf{Observed Behavior}
To address the issues from the first iteration, we added a directive restricting the response format to a JSON object only. This modification successfully constrained the output, making it more reliable and compatible with the simulation. However, while some flocking behavior emerged, it was inconsistent. Small clusters formed briefly, but overall alignment and coherence were weaker than expected.

\item \textbf{Analysis}
We compared the resulting headings with those produced by a rule-based model. While some calculated headings were numerically similar, they often pointed in opposite directions. This suggested ambiguity in the LLM's coordinate system. Since NetLogo employs a compass convention for heading calculations, we decided to explicitly specify this convention in the next iteration.
\end{itemize}

\textbf{Iteration 3}
\begin{itemize}
\item \textbf{Prompt Text}

An instruction was added to specify that the compass convention should be used.

\begin{lstlisting}[language=json,firstnumber=1,basicstyle=\small]
You are an agent in a 2D simulation. Following the compass convention, your task is to determine your new heading based on the flocking principles of separation turn, alignment turn (average heading of neighbors), and coherence turn (average heading towards flockmates)
...
\end{lstlisting}

\item \textbf{Observed Behavior}
By explicitly specifying the compass convention for heading calculations, flocking behavior improved. Larger clusters formed compared to previous iterations. However, flocking remained inconsistent, as some agents moved in seemingly random directions.

\item \textbf{Analysis}
Examining the erratic headings, we requested the LLM to explain its calculations. When generating a reasoning process before outputting the final heading, the LLM produced correct answers. However, errors occurred when providing only the numerical result. This highlighted the need for a structured reasoning process, or ``chain of thought,'' to ensure accurate heading calculations.
\end{itemize}

\textbf{Iteration 4}
\begin{itemize}
\item \textbf{Prompt Text}

A new key, ``rationale'', was added to the JSON output to encourage reasoning before determining the final heading.

\begin{lstlisting}[language=json,firstnumber=1,basicstyle=\small]
...
Provide your final new heading after applying these rules, expressed as an angle in degrees. The result should be in JSON format only, with the keys and values: "rationale" (value: your explanation) and "new-heading" (value: heading in degrees).
...
\end{lstlisting}

\item \textbf{Observed Behavior}
Introducing the ``rationale'' key significantly improved flocking behavior. The agents demonstrated more consistent heading adjustments, enhancing the emergence of flocking dynamics. However, occasional errors persisted, particularly when agents needed to turn counterclockwise to reach a nearby target heading.

\item \textbf{Analysis}
The ``rationale'' key enabled the LLM to engage in a structured thought process, substantially improving flocking behavior. However, some agents still moved in the opposite direction when making small adjustments, particularly for counterclockwise turns. This suggested that additional guidance was necessary to ensure agents always chose the shortest rotation path.
\end{itemize}

\textbf{Iteration 5}
\begin{itemize}
\item \textbf{Prompt Text}

An explicit instruction was added to ensure the shortest rotational path (clockwise or counterclockwise) was always chosen when adjusting the heading.

\begin{lstlisting}[language=json,firstnumber=1,basicstyle=\small]
You are an agent in a 2D simulation. Following the compass convention, your task is to determine your new heading based on the flocking principles of separation turn, alignment turn (average heading of neighbors), and coherence turn (average heading towards flockmates). The parameters for these principles are: maximum-separate-turn, maximum-align-turn, maximum-cohere-turn, minimum-separation-distance. The simulation provides the following information: Current heading, Neighbors in vision radius.  When calculating the alignment turn, always choose the shortest path (clockwise or counterclockwise) to align with the average heading of neighbors.
...
\end{lstlisting}

\item \textbf{Observed Behavior}
By explicitly instructing the model to select the shortest path to the target heading, flocking behavior improved significantly. The LLM-driven agents formed larger, more stable flocking clusters, achieving performance comparable to the original, rule-based NetLogo model.

\item \textbf{Analysis}
Including the shortest-path directive ensured that LLM-based agents correctly aligned their heading adjustments with both LLM-based and rule-based agents. This modification resolved the previously observed issues, leading to a more coherent and emergent flocking behavior.
\end{itemize}

\subsubsection{Prompt Deployment}

This task uses the same prompt deployment mechanism as Experiment~1. Communication with the LLM is handled via the OpenAI API using the chat.completions mechanism, which supports context-free messaging. A system prompt defines the task and rules, followed by a user prompt providing real-time environmental data.

At each simulation step, NetLogo translates the agent's perception into the input variables in the prompt, including the heading and position of other agents within its vision radius, into the input variables used in the prompt. This ensures the LLM has an accurate, up-to-date view of the environment. The LLM then generates a response formatted in JSON, specifying the agent's actions. The following example prompt illustrates this process:

\begin{itemize}
\item \textbf{System Prompt}
\begin{lstlisting}[language=json,firstnumber=1,basicstyle=\small]
You are an agent in a 2D simulation. Following the compass convention, your task is to determine your new heading based on the flocking principles of separation turn, alignment turn (average heading of neighbors), and coherence turn (average heading towards flockmates). The parameters for these principles are: maximum-separate-turn, maximum-align-turn, maximum-cohere-turn, minimum-separation-distance. The simulation provides the following information: Current heading, Neighbors in vision radius. When calculating the alignment turn, always choose the shortest path (clockwise or counterclockwise) to align with the average heading of neighbors. 

Provide your final new heading after applying these rules, expressed as an angle in degrees. The result should be in JSON format only, with the keys and values: 'rationale' (value: your explanation) and 'new-heading' (value: heading in degrees).
\end{lstlisting}

\item \textbf{Possible User Prompt}
\begin{lstlisting}[language=json,firstnumber=1,basicstyle=\small]
These are the flocking parameters: 
   -Maximum separate turn: 1.5, 
   -Maximum align turn: 5, 
   -Maximum cohere turn: 3, 
   -Minimum separation: 1.5; 
   
This is your current environment: 
   -Current heading: 138 deg, 
   -Neighbors in vision radius: neighbor_1: x: 0.53, y: -3.69, heading: 248 deg;
\end{lstlisting}

\item \textbf{Possible Response of the LLM}
\begin{lstlisting}[language=json,firstnumber=1,breaklines=true]
{
  "rationale": Given the current heading of 138 degrees and one neighbor at (0.53, -3.69) with a heading of 248 degrees, we first calculate the separation turn. The distance to the neighbor is sqrt(0.53^2 + (-3.69)^2) = 3.73, which is greater than the minimum separation distance of 1.5, so no separation turn is needed. For alignment, the average heading of the neighbor is 248 degrees. The shortest path to align with 248 degrees from 138 degrees is clockwise, which is a turn of 110 degrees. However, the maximum align turn is 5 degrees, so we turn 5 degrees clockwise to 143 degrees. For coherence, since there is only one neighbor, the average heading towards the neighbor is the same as the neighbor's heading, 248 degrees. The shortest path to align with 248 degrees from 143 degrees is clockwise, which is a turn of 105 degrees. However, the maximum cohere turn is 3 degrees, so we turn 3 degrees clockwise to 146 degrees. The final new heading is 146 degrees.,
  "new-heading": 146
}
\end{lstlisting}
\end{itemize}

\subsection{Monitoring Behavior}

Key metrics include the degree of separation maintained between agents, the consistency of alignment with the average heading of nearby birds, and the effectiveness of cohesion in moving toward the flock's center of mass. The observed behaviors and performance analyses are documented in the results section.

\section{Results}\label{sec:results}
\subsection{Experiment 1: Ant Foraging with Structured Rule-Based Prompting}\label{subsec:experiment_1}


The following three variants of the ants foraging simulation were applied:
\begin{enumerate}
    \item The original NetLogo model (henceforth simply called ``NetLogo'').
    \item The model in which the rule-governed ants of the original model are replaced by LLM-governed ants (henceforth called ``LLM'').
    \item A hybrid model in which half of the ants are rule-governed and the other half are LLM-governed (henceforth called ``Hybrid'').
\end{enumerate} 

In all simulations, we used a colony of 10 ants, three food patches to be exploited, and a stopping criterion of 1000 simulation steps. Moreover, each experiment was repeated five times (with different seeds). The efficacy of each model was assessed by quantifying the total amount of food gathered within these 1000 simulation steps. 

\subsubsection{Food collection behavior}\label{subsubsec:food_collection}
\begin{figure}[ht]
    \centering
    \includegraphics[width=0.8\linewidth]{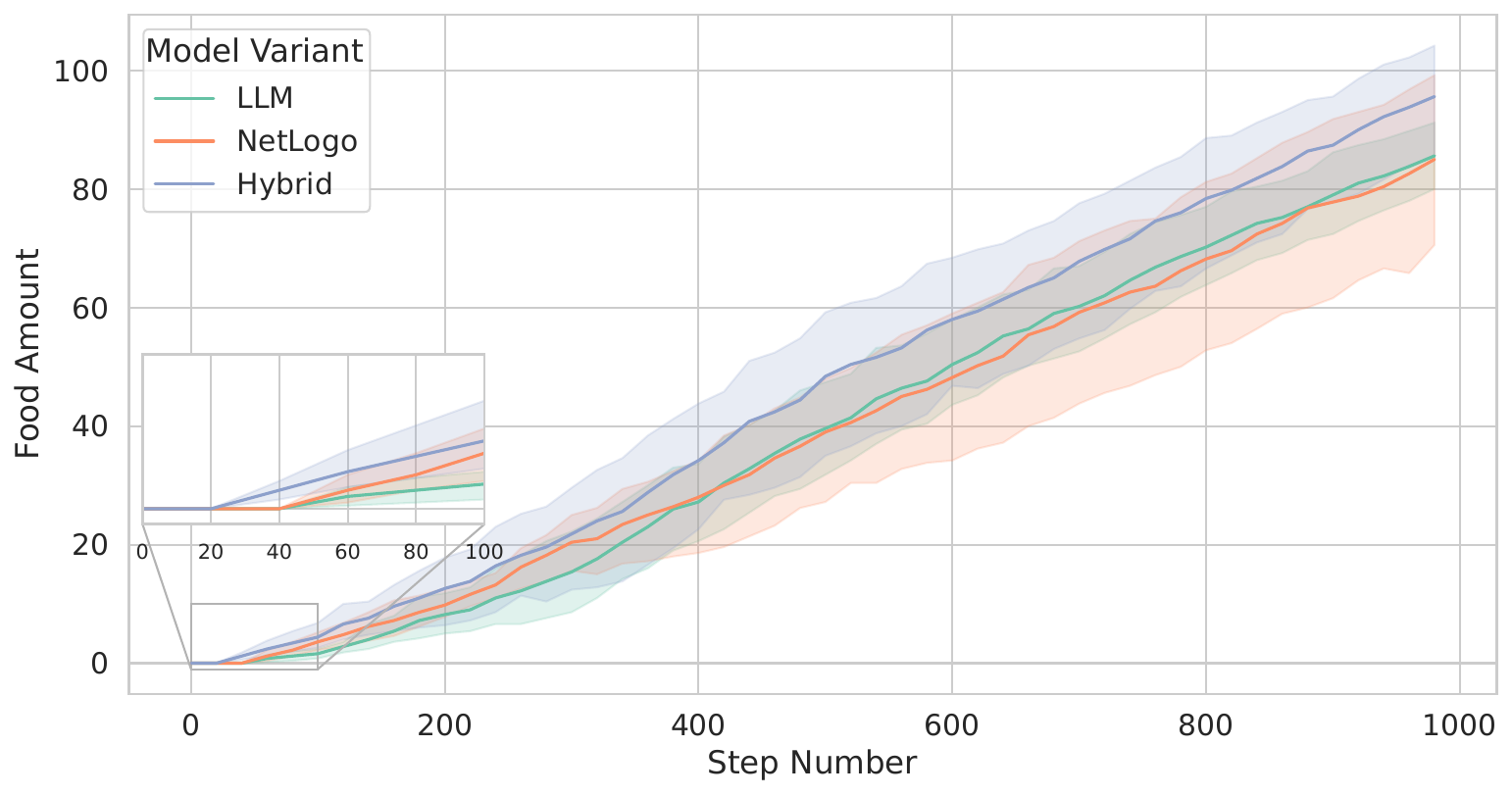}
    \caption{Comparison of the total food collected across the three tested models: NetLogo (represented by the orange line), LLM (green line), and Hybrid (blue line). This visualization highlights the food collection performance differences among the different models over five runs with different seeds. The lines represent the means, while the shaded areas indicate the standard deviations.}
    \label{fig:ant_colony_food_amount}
\end{figure}

\autoref{fig:ant_colony_food_amount} shows the total amount of food collected over 1000 simulation steps for the three different model variants. NetLogo and LLM perform similarly in terms of the ants' success in bringing food back to the nest, with both models accumulating approximately 85 units of food by the end of the simulation. However, the standard deviation for NetLogo is around 20, whereas LLM displays a much lower standard deviation of about 7. 

In contrast, the Hybrid model outperforms the other two variants, collecting an average of approximately $95$ units of food with a standard deviation of about $12$. This superior performance is due to the combination of the behavioural differences between LLM-guided ants and rule-based ants. The zoomed inset in~\autoref{fig:ant_colony_food_amount}, for example, shows that Hybrid starts returning food to the nest at around 20 simulation steps, whereas LLM and NetLogo begin this process at about 40 steps. This means that, for some reason, the Hybrid variant is more efficient in quickly finding food sources.

\begin{figure}[h]
    \centering
    \includegraphics[width=0.8\linewidth]{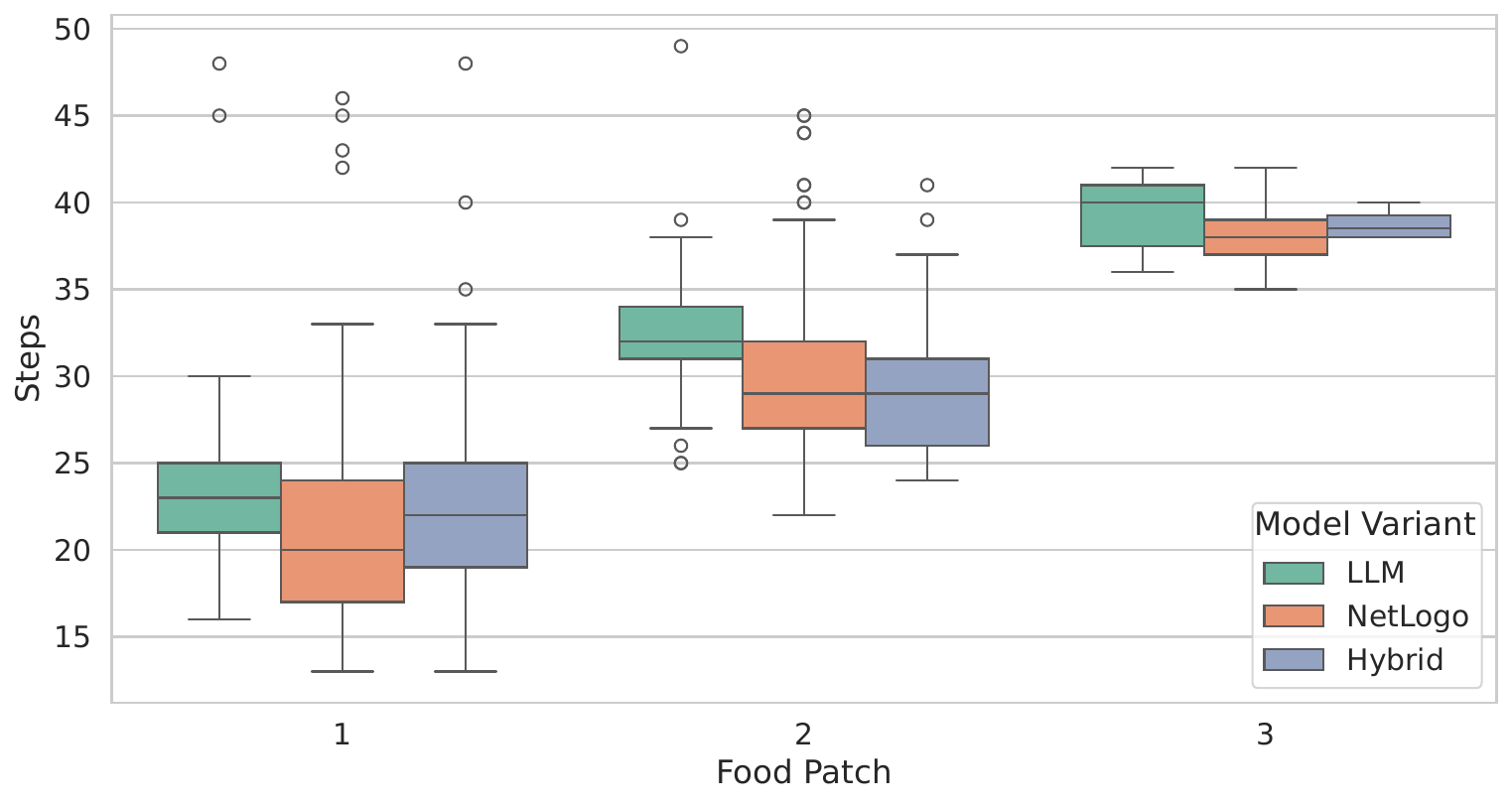}
    \caption{The average number of steps taken by an ant to return to its nest after picking up food (for food patches 1--3). The green boxplots represent the simulations of LLM, the orange boxplots those of NetLogo, while the blue boxplots show the results of Hybrid. Each boxplot spans from the first to the third quartile, with the vertical line within the box indicating the median. The whiskers extend to represent the minimum and maximum number of steps taken, while the circles denote outliers.}
    \label{fig:steps_return_food}
\end{figure}

\begin{table}
\centering
\caption{Statistics concerning the average amount of steps taken by an ant to return food to the nest.}
\label{tab:food_return}
\begin{tabular}{c|l|ccccccc}
\toprule
 \textbf{Food Patch} &  \textbf{Model Variant} & \textbf{Mean} &  \textbf{Std} & \textbf{Min} & 20\% & 50\% & 75\% & \textbf{Max} \\ \midrule    
& LLM 	& 23.04 & 3.34 &	16.0 & 21.0 & 23.0 & 25.0 & 48.0 \\
1 & NetLogo & 21.0 & 5.3 & 13.0 & 17.0 & 20.0 & 24.0  & 46.0\\
& Hybrid & 21.98 &  4.32 & 13.0 & 19.0 	& 22.0 & 25.00 & 48.0 \\ \midrule
& LLM & 32.3 & 3.41 & 25.0 & 31.0 & 32.0 & 34.0 & 49.0\\
2 & NetLogo & 30.16 & 4.93 & 22.0 & 27.0 & 29.0 & 32.0 & 45.0\\
& Hybrid & 29.46& 3.90 & 24.0 & 26.0 & 29.0 & 31.00 & 41.0\\ \midrule
& LLM & 39.29 & 2.36 & 36.0 & 37.5 & 40.0 & 41.0 & 42.0\\
3 & NetLogo & 38.11 & 2.02 & 35.0 & 37.0 & 38.0 & 39.0 & 42.0 \\
& Hybrid & 38.75 & 0.96 & 38.0 & 38.0 & 38.5 & 39.25 & 40.0\\ \bottomrule
\end{tabular}
\end{table} 

The average number of simulation steps taken by an ant to return to its nest after picking up food is depicted as boxplots in~\autoref{fig:steps_return_food}. This plot illustrates the effectiveness of the three model variants concerning individual ants. Generally, ants in NetLogo (rule-governed ants) require fewer steps than those controlled by the LLM. The LLM-guided ants demonstrate consistent foraging behavior across the different experiments, particularly for food patches 1 and 2. Notably, food patch 1 is the closest to the nest, while food patch 3 is the farthest away. 
Detailed statistics, including the three quartiles, mean, standard deviation, and minimum and maximum amounts of food collected, are provided in~\autoref{tab:food_return}.

\begin{figure}
    \centering
    \includegraphics[width=0.8\linewidth]{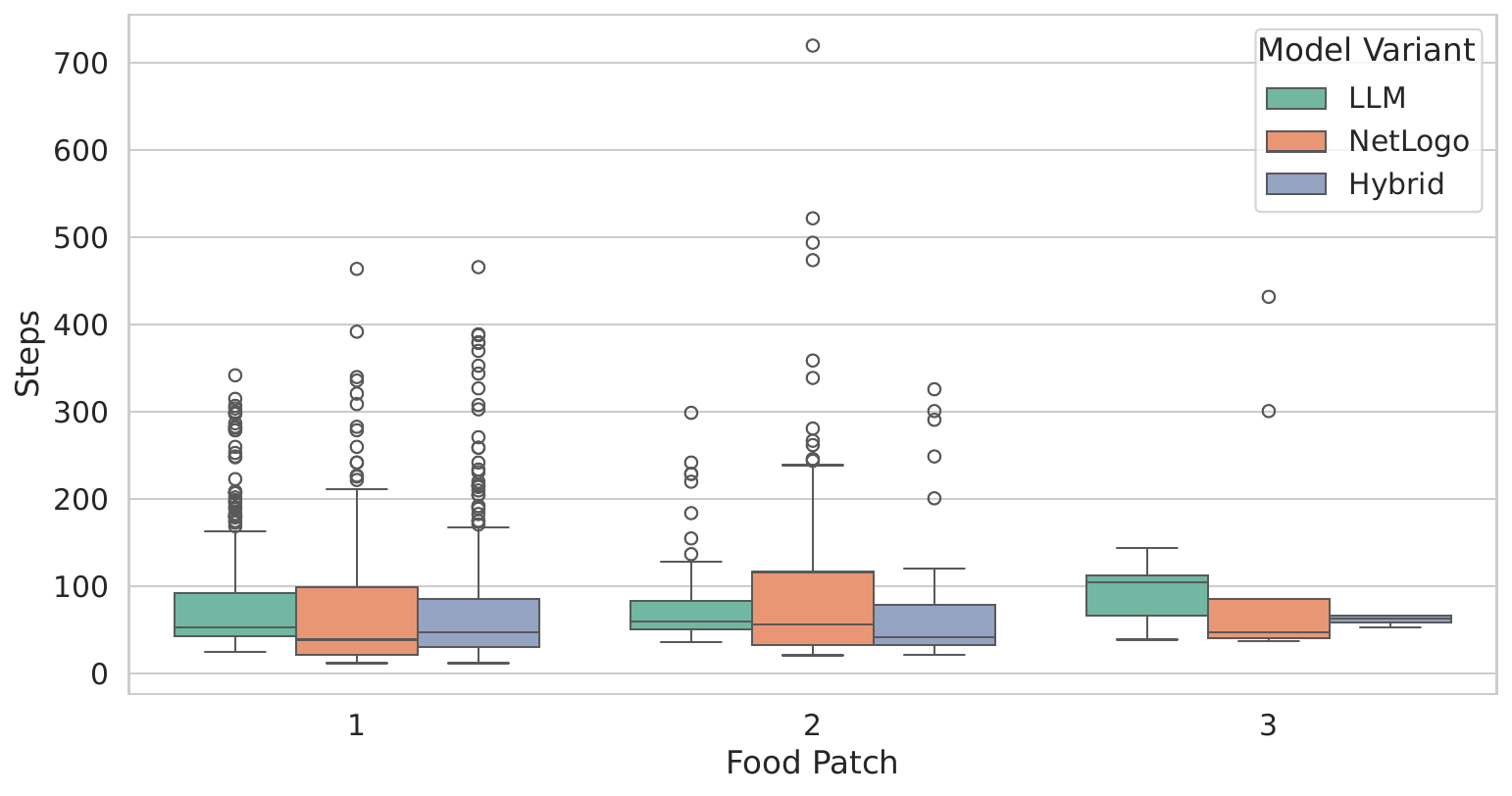}
    \caption{Average number of steps taken by an ant from leaving the nest to finding a food source. Each boxplot spans from the first to the third quartile, with the vertical line within the box indicating the median. The whiskers extend to represent the minimum and maximum number of steps taken, while the circles denote outliers.}
    \label{fig:steps_search_food}
\end{figure}

The average number of steps taken by an ant from leaving the nest until finding a food source is represented in~\autoref{fig:steps_search_food}. We specifically track and count ants that are not carrying food and are exploring their environment until they start to carry the food. Hybrid demonstrates consistent performance in finding food patches 1 and 2. In contrast, LLM and NetLogo display a more variable behavior during food searches. Notably, concerning food patch 1, the models exhibit a higher number of outliers, which can be attributed to the ants' initial exploration of the environment before encountering the food. A notable outlier is observed in the context of NetLogo and food patch~2, where an ant required $720$ steps to find food. 
Detailed statistics are listed in~\autoref{tab:food_search}.

\begin{table}
\centering
\caption{Statistics for the average amount of steps taken by an ant to find and collect food.}
\label{tab:food_search}
\begin{tabular}{c|l|ccccccc}
\toprule
 \textbf{Food Patch} &  \textbf{Model Variant} & \textbf{Mean} &  \textbf{Std} & \textbf{Min} & 20\% & 50\% & 75\% & \textbf{Max} \\ \midrule    
& LLM 	& 79.65 & 63.25 & 25.0  & 43.0 & 53.0 & 92.0 & 342.0 \\
1 & NetLogo & 71.48 & 72.77 & 12.0 & 22.0 & 39.0 & 99.0  & 464.0\\
& Hybrid & 71.42 & 68.88 & 12.0 & 31.0 & 47.0 & 86.0 & 466.0 \\ \midrule
& LLM & 79.44 & 50.04 & 36.0 & 51.0 & 60.0 &	83.0 & 299.0 \\
2 & NetLogo & 93.74 & 102.09 & 21.0 & 33.0 & 56.0 & 116.50 & 720.0\\
& Hybrid & 73.81 & 74.81 & 22.00 & 32.75 & 42.00 & 78.50 & 326.0\\ \midrule
& LLM & 92.29 & 36.53 & 39.0 & 66.50 & 105.0 & 112.50 & 144.0\\
3 & NetLogo & 123.33 & 142.92 & 37.0 & 41.0 & 47.0 & 86.0 & 432.0 \\
& Hybrid & 61.25 & 6.18 & 53.0 & 58.25 & 63.0 & 66.0 & 66.0\\ \bottomrule 
\end{tabular}
\end{table} 

\subsection{Experiment 2: Bird Flocking Simulation with Knowledge-driven Prompts}
\label{experiment_2}

The following two model variants were experimentally tested and evaluated:
\begin{enumerate}
    \item The original NetLogo model (henceforth simply called ``NetLogo'', just like in the ant foraging case).
    \item The model in which some of the rule-governed birds of the original model are replaced by LLM-governed birds (henceforth called ``Hybrid'').
\end{enumerate} 

\begin{figure}[t]
    \centering
    \includegraphics[width=0.7\linewidth]{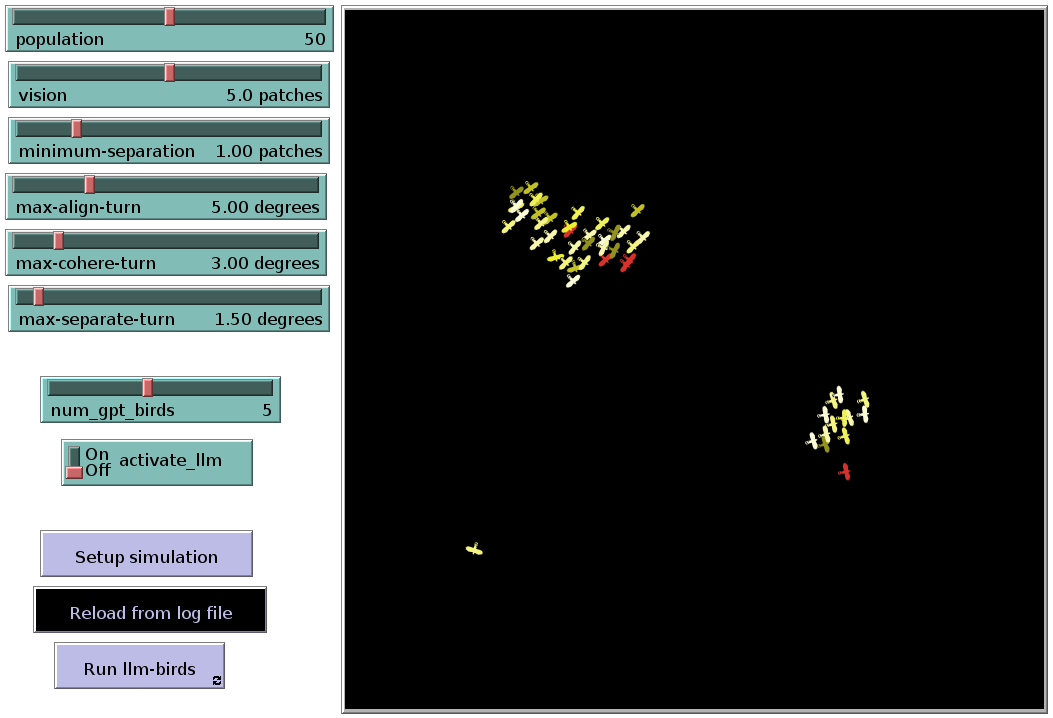}
    \caption{Bird flocking simulation in the NetLogo environment: yellow birds follow rule-based behavior, while red birds are guided by the LLM.}
    \label{fig:bird_flocking_interface}
\end{figure}

In all simulations, we used a flock of 30 birds and a simulation length of 800 steps. In the case of Hybrid, five of 30 rule-based birds are replaced by LLM-guided birds. Moreover, each experiment was repeated five times (with different seeds). The effectiveness of the flocking behavior is evaluated by measuring the distances and angular disparities between birds across the entire simulation. Figure~\ref{fig:bird_flocking_interface} depicts the flocking simulation executed in the NetLogo environment, featuring a heterogeneous population of 25 rule-based and five LLM-guided birds.

\subsubsection{Flocking behavior}\label{subsubsec:flocking_behavior}
\begin{figure}[h]
    \centering
    \includegraphics[width=0.8\linewidth]{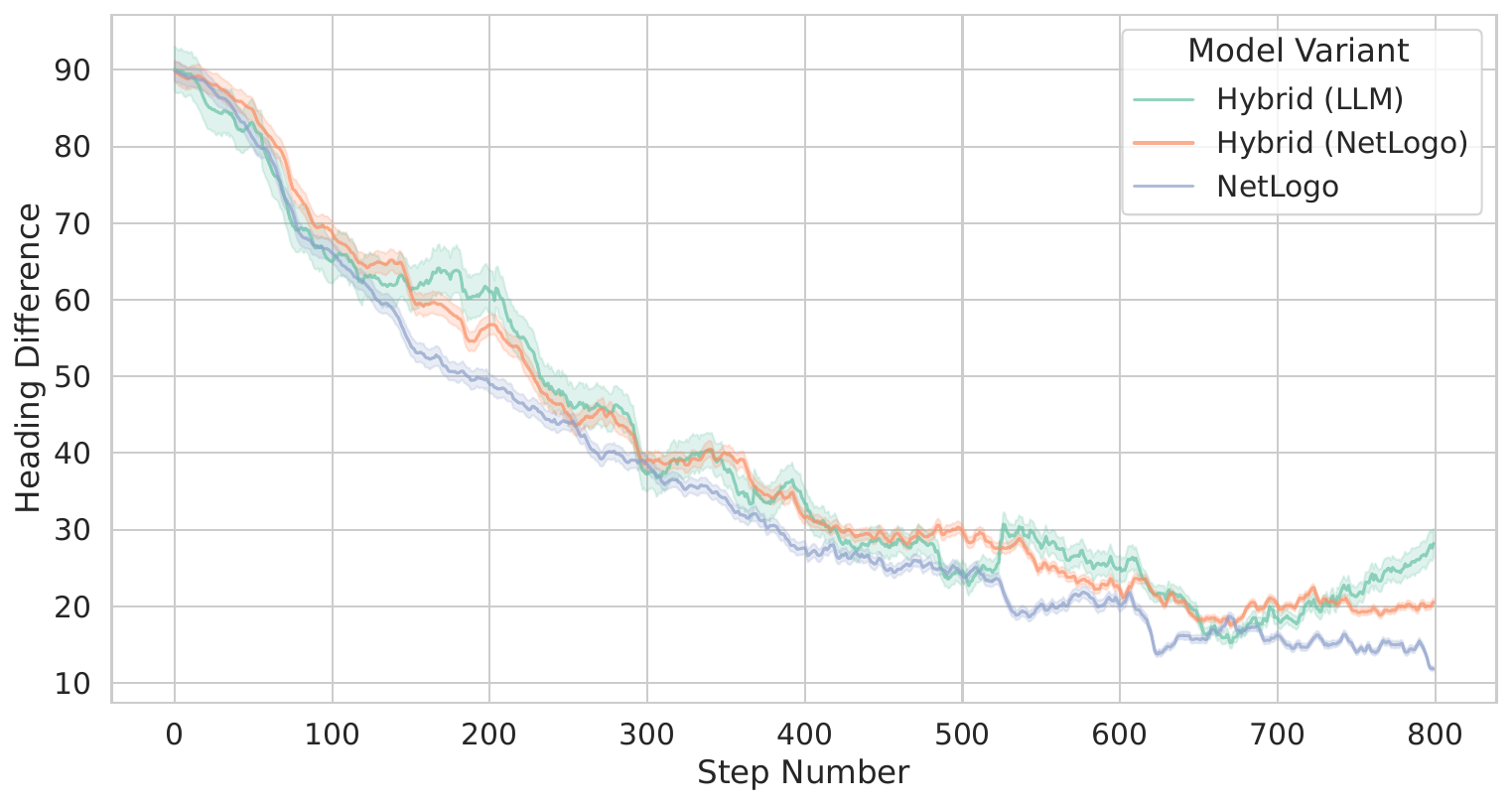}
    \caption{Comparison of bird flocking heading differences across two simulation approaches: original NetLogo (blue line) and Hybrid (orange and green lines). In fact, the orange line shows the behavior of the 25 rule-based birds of Hybrid, while the green line presents the behavior of the 5 LLM-guided birds of Hybrid. The lines represent the means, while the shaded areas indicate the standard deviations.}
    \label{fig:heading_differences}
\end{figure}
\begin{figure}[h]
    \centering
    \includegraphics[width=0.8\linewidth]{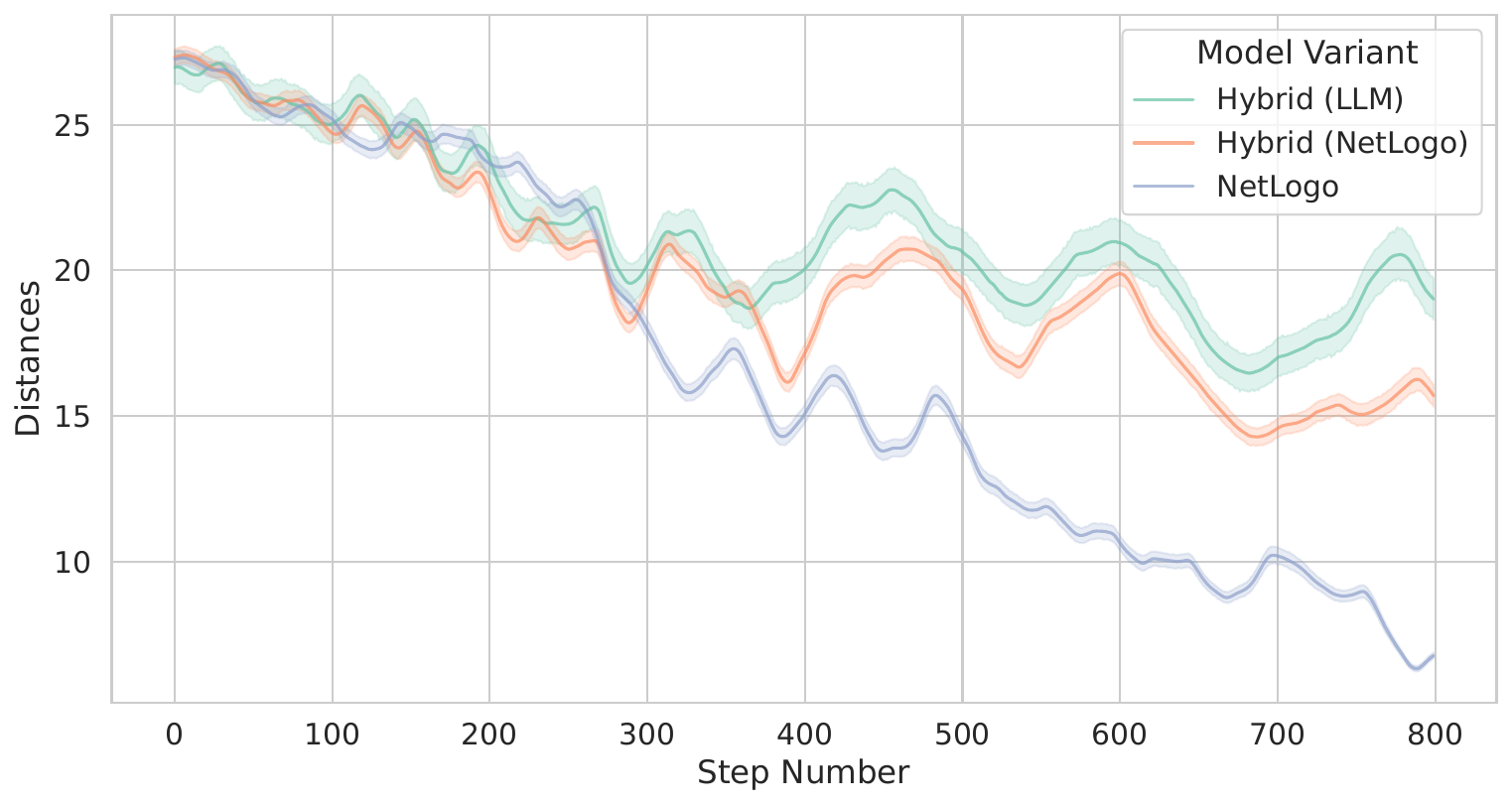}
    \caption{Comparison of average bird distances across the two tested model variants}
    \label{fig:distances_hybrid_rule-based}
\end{figure}

\autoref{fig:heading_differences} compares the differences in the birds' heading directions between two model variants, as outlined above. However, note that the heading differences between the rule-based birds and all other birds in model variant Hybrid (orange line) are separated from the heading differences between the LLM-guided birds and all other birds of Hybrid (green line). 

The results shown in \autoref{fig:heading_differences} allow to make the following observations. While the two bird types of Hybrid show a similar evolution of the heading differences, the rule-based birds of the original NetLogo model show somewhat lower heading differences. We anticipate that with longer simulation runs, the heading differences of the two model variants would converge to similar values. We also observed that the LLM-guided birds tend to congregate at the outer peripheries of the flocks, positioning themselves further away from the flocks' center. An example of this behavior is visualized in~\autoref{fig:bird_flocking_interface} (see the flock on the right) and also illustrated in~\autoref{fig:distances_hybrid_rule-based} which indicates the average distances between birds. We hypothesize that this rather ``conservative'' behavior of the LLM-guided birds contributes to greater heading differences among the rule-based birds of Hybrid, as this behavior introduces slight perturbations in the flocking dynamics. Another possible interpretation involves the internal representation of distance within the LLM. Although we define distance in Euclidean space and provide these distances as float values to the language model, it may interpret and represent distances in a different manner.

\begin{figure}
    \centering
    \includegraphics[width=0.8\linewidth]{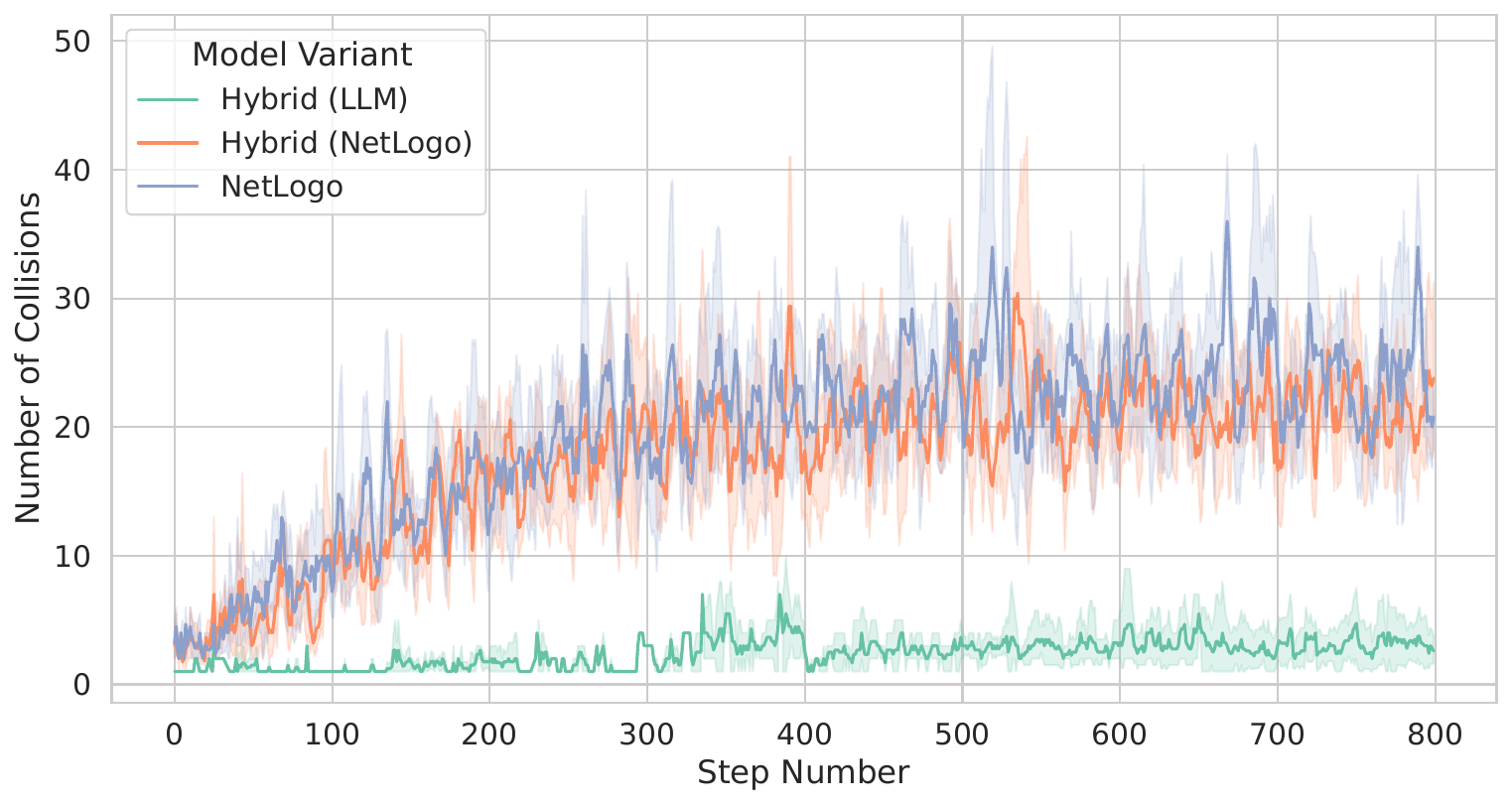}
    \caption{Collisions between birds. A collision occurs when the distance $d$ between birds is at most one (that is, $d \leq 1$).}
    \label{fig:collisions_birds}
\end{figure}

We further investigated the behavior of rather staying at the border of flocks by examining collisions between birds, which are defined as occasions in which the Euclidean distance between two birds is smaller than one. In fact, it turns out that, throughout a simulation, the LLM-guided birds try to avoid collisions; see~\autoref{fig:collisions_birds}. In contrast, the rule-based birds from Hybrid and those from NetLogo, exhibit a much higher number of collisions.

\begin{figure}
    \centering
    \includegraphics[width=0.9\linewidth]{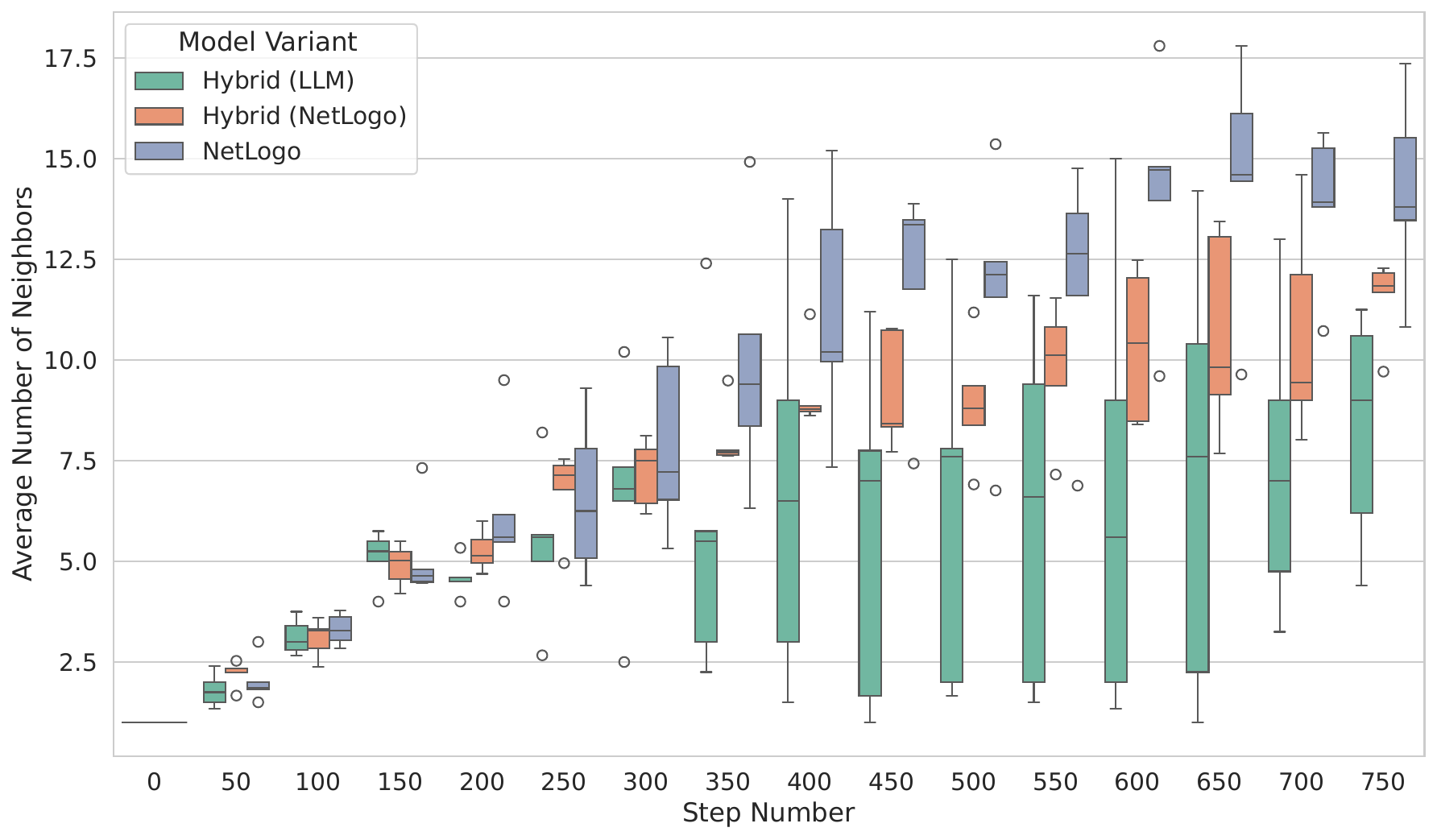}
    \caption{Average number of neighbors: A neighbor is defined as any entity within a distance $d$ such that $1 \leq d \leq 5$, thereby excluding collisions. Furthermore, we establish a heading difference criterion of $h\leq 15$.}
    \label{fig:average_number_neighbours}
\end{figure}
\begin{table}
\centering
\caption{Statistics for the average number of flocking neighbors. The values are aggregated over all steps and experiments.}
\label{tab:flocking_statistics}
\begin{tabular}{l|cccccccc}
\toprule
\textbf{Model Variant} & \textbf{Mean} & \textbf{Median} & \textbf{Std} & \textbf{Min} & 20\% & 50\% & 75\% & \textbf{Max} \\ \midrule
Hybrid (LLM)  & 6.27 & 5.40 & 4.23 & 1.00 & 2.75 & 5.40 & 8.63 & 17.80 \\ 
Hybrid (NetLogo) & 9.23 & 9.44 & 4.30 & 1.04 & 6.44 & 9.44 & 12.25 & 16.76 \\ 
NetLogo &11.42 & 11.24 & 6.27 & 1.04 & 6.02 & 11.24 & 16.87 & 22.56 \\ \bottomrule
\end{tabular}
\end{table}

Furthermore, triggered by our earlier observations, we examined the average number of neighbors of a bird, as shown in~\autoref{fig:average_number_neighbours}. Hereby, we define two birds as neighbors if they are at a distance greater than one (no collision) and within a distance $d$ of at most five (that is, $1 < d \leq 5$). Moreover, we require a heading difference of $h \leq 15$. As expected, rule-based birds exhibit the highest number of neighbors, while the LLM-guided birds display the lowest number, a result of their conservative behavior. Statistics on the average number of flocking neighbors can be found in~\autoref{tab:flocking_statistics}. 

\section{Discussion}

We explored two complementary strategies for integrating Large Language Models (LLMs) into multi-agent simulations: (i) a structured, rule-based approach in an ant foraging context, and (ii) a principle-based, knowledge-driven approach in a bird flocking scenario. In both cases, our experiments demonstrated how LLMs can support swarm-like behaviors: guiding ants to locate and retrieve food by following pheromone trails, and prompting “birds” to coordinate alignment according to core flocking principles. Overall, the LLM-driven agents performed comparably to their fully rule-based counterparts, but they sometimes displayed notable differences in how they interpreted and prioritized local cues when relying on text-based decision-making.

A key theme across both simulations was the importance of iterative prompt-tuning, which proved essential for producing consistent and context-appropriate responses. In the ant foraging simulations, early prompts did not specify what ants should do if no pheromone or nest scent was present, leading to confusion or inaction. Through multiple rounds of tuning, we added directives, such as ``move away from the nest when no pheromone signals are detected'' that encouraged exploration. Similarly, clarifying that nest scent should take precedence over pheromone while carrying food helped ants more reliably locate and deposit resources. Following these refinements, the foraging performance of the LLM-driven ants nearly matched that of the standard NetLogo model.

A new insight arose from the hybrid simulations, in which a portion of the ant colony was rule-based while the rest was LLM-driven. These mixed colonies often outperformed both purely rule-based and purely LLM-based groups. One possible explanation is that deterministic if-then logic efficiently manages well-understood aspects of foraging, while LLM-driven exploration provides adaptability in more uncertain situations. Thus, combining traditional rules with text-based reasoning can yield more robust foraging strategies. However, this seemingly better performance of the hybrid populations warrants further investigation. We recommend running the simulations for longer durations so that the colony has enough time to collect any remaining pieces of food, which may help clarify the mechanisms driving this performance advantage.

In the bird flocking simulations, using longer prompts that highlighted alignment, separation, and cohesion improved stability. Early prompts did not define heading conventions (e.g., $0^\circ$ = north, $90^\circ$ = east), causing erratic turns and reversals. After establishing the conventions and clarifying the short-turn logic (which favored minimal angular adjustments), the flocks became more cohesive. However, LLM-driven birds generally stayed slightly farther from the flock center and experienced fewer collisions than their rule-based counterparts, indicating that LLMs can interpret spatial cues in subtly different ways while still maintaining coherent swarm behavior.

In both scenarios, we observed that LLM decision-making can function effectively in a ``stateless'' manner, relying on complete contextual details at every step. This guarantees that the model consistently acts on relevant information but also necessitates highly detailed prompts. Failing to include key details---like pheromone intensity or heading conventions---can result in ambiguous or incorrect actions. Expanding this approach to incorporate short-term memory or more sophisticated environmental representations could enable LLM-driven agents to maintain internal states that more closely resemble those in traditional agent-based models.

Together, these results confirm that LLMs can serve as flexible engines for agent behaviors that align with swarm principles, offering adaptive, context-driven responses. They also highlight how prompt design and iterative refinement are central to achieving the desired outcomes. Even small changes in the prompts, such as specifying the angle to rotate or how to handle conflicting signals, can significantly influence emergent group-level patterns. This underscores both the potential and the complexity of integrating LLMs into agent-based simulations, where subtle details of agent logic can greatly affect collective behavior.

Finally, regarding potential drawbacks of our approach, two key issues must be noted: computation time and cost. First, the interaction between an agent (such as an ant or bird) and the remote LLM at each iteration of a simulation requires significantly more computation time compared to executing simple rules within NetLogo. Second, utilizing GPT-4o through an API incurs token-based costs associated with the use of an API key. However, this rise in computation time and cost could be reduced by deploying a smaller, locally hosted model after training and fine-tuning. 

\section{Conclusion}

By applying LLMs to two classic multiagent models, ant foraging and bird flocking, this study shows that LLMs can serve as a viable alternative or complement to traditional rule-based logic in achieving effective swarm-like dynamics.

In ant foraging simulations, LLM-guided ants gathered food at rates comparable to ants of the standard NetLogo model, as long as their instructions were meticulously designed. Moreover, hybrid colonies that integrated LLM-driven and rule-based ants outperformed uniform groups, indicating that the combination of deterministic efficiency and text-based reasoning can be mutually beneficial.

In bird flocking, LLM-driven agents adhered to the separation, alignment, and cohesion principles to form cohesive flocks. While heading convergence sometimes lagged behind purely rule-based simulations, the resulting formations remained visually coherent. Notably, LLM-based birds adopted slightly more peripheral positions, indicating that nuanced differences in textual instructions such as how distance and turning are interpreted, can shape global flock patterns.

These experiments emphasize the crucial role of iterative prompt tuning in aligning LLMs with specific multi-agent objectives. Meticulous attention to prompt length, structure, and content is necessary to ensure reliable, context-aware behavior at each time step. At the same time, this reliance on well-tuned prompts opens up exciting avenues for further research: more complex simulations might benefit from greater LLM-driven adaptability, especially if additional mechanisms like partial memory or reinforcement signals are introduced to move beyond purely stateless approaches. Ultimately, this work underscores the potential for advanced language models, guided by carefully designed prompts, to enrich or even extend the capabilities of traditional agent-based models, offering new perspectives on swarm intelligence, self-organization, and emergent behaviors.

\section*{Acknowledgements}

This research was supported by the EUTOPIA Science and Innovation Fellowship Programme and funded by the European Union Horizon 2020 programme under the Marie Sklodowska-Curie grant agreement No 945380.

Christian Blum was supported by grant PID2022-136787NB-I00 funded by MCIN/AEI/10.13039/501100011033. 

\section*{Disclaimer}

This article reflects only the author's view and the EU Research Executive Agency is not responsible for any use that may be made of the information it contains.

\bibliographystyle{unsrtnat}






\end{document}